# Tilted Nanofiber Bragg Grating as an efficient photon collector from single quantum emitters: A new avenue

SUBRAT SAHU & RAJAN JHA*

*Nanophotonics and Plasmonics Laboratory, School of Basic Sciences, Indian Institute of Technology Bhubaneswar, 752050, Odisha, India*

**We report a new gateway towards the light-matter interaction of quantum light emitter's spontaneous emission in optical nanofiber-based on nanocavities tilted by some angle with respect to the plane of the fiber cross-section. Our result shows that the Purcell factor of a quantum light emitter inside the nanofiber is nine times more than that of an emitter on the cavity's surface. A light emitter's coupling efficiency into the tilted grating structure's guided mode is ~ 65%, with 70 grating periods. The tilted angle can be optimized to get more coupling efficiency with a decrease in the Purcell factor. This tilted structure has a high Q factor ~ 1408.71 and low mode volume ~ 0.66 μm$^3$. For the quantum information and quantum photonics application, this system attracts researchers to a new direction in the quantum world.**

*OCIS codes:* (060.3735) Fiber Bragg grating; (060.5565) Quantum Communications; (020.5580) Quantum electrodynamics; (270.1670) Coherent optical effects

Efficient and robust light-matter interaction plays a crucial role in the quantum-based system to control the coupling of photons emitted from the quantum emitters (QEs) for a variety of applications [1–3]. Optical nanofiber (ONF) emerges as an efficient and configurable interface/workbench for such QEs to study light-matter interaction [1] due to its sub-wavelength nature and tight confinement of the high-intensity transverse mode around the fiber. The photons emitted from the QEs placed over the waist region of the ONF can be precisely coupled to the nanofiber's guided mode via a strong evanescent field and to the free space through radiated mode. In the recent past, the coupling of light emitters with the ONF had experimentally been investigated using semiconductor quantum dots [4], neutral atoms [5], molecules [6], nitrogen-vacancy (NV) in diamond [7,8], and 2D materials [9]. Maximum coupling efficiency of 30% has been achieved by placing an NV center on the surface of ONF [10]. However, theoretically, a trapped atom can absorb 8% of the nanofiber's guided light placed about a distance of 200 nm away from the nanofiber, and it decreases with the increase in distance between the atom and nanofiber [10]. Experimentally 20 % of photons from the single QE have been coupled to the nanofiber's guided mode by precisely positioning the QEs on the surface of ONF [11].

To further increase the coupling enhancement, systems like micropillar cavity using distributed Bragg reflector (DBR) [12], photonic crystal (PhC) cavity [13], and fiber Bragg grating (FBG) [14] have been used. Hakuta *et al.* reported nanofiber-based holes using the femtosecond laser ablation method and focused ion beam (FIB) milling technique [14,15]. Takeuchi *et al.* reported the enhancement of emitted photon from the quantum dot coupled with the nanofiber Bragg cavity (NFBC) [16]. Nayak *et al.* reported an array of nano crater on an optical nanofiber using the femtosecond laser ablation technique to show high transmission of 87% [17]. Keloth *et al.* create a centimeter-long nanofiber cavity for strong-coupling cavity quantum electrodynamics (cQED) [18]. Yella *et al.* formed PhC cavities using an external grating to enhance spontaneous emission from the single quantum dot guided into the optical nanofiber [15,19]. Chromaic *et al.* proposed a nanofiber cavity structure combination of FBG and PhC cavity to make a resonant cavity [20,21]. However, one of the most matured and commercially available fiber systems like Tilted Fiber Bragg Grating (TFBG) has not to be explored. TFBG fabrication uses laser writing on a doped fiber by an interference pattern. This flexible approach allows changing of the periods as well as tilt angles conveniently. For a large number of the identical grating, the interference pattern is generated by a diffractive phase mask located close to the fiber [22]; tilt angle is monitored by rotating the phase mask and fiber, and the resonance condition for TFBG is given as [23],

$$\lambda_r = (n_{eff}^{core}(\lambda_r) + n_{eff}^{r}(\lambda_r))\Lambda / \cos(\theta) \quad (1)$$

Here $\lambda_r$ is Bragg resonance wavelength, $n_{eff}^{core}$ is the effective index of core mode, $n_{eff}^r$ is the cladding effective index, $\Lambda$ is the grating period, and $\theta$ is the tilt angle of the grating planes with respect to the plane of the fiber cross-section.

In this letter, we have reported tilted grating in an optical nanofiber called tilted nanofiber Bragg grating (TNFBG) as an efficient photon collector to enhance QE's spontaneous decay. We found that TNFBGs have high-quality factors (Q), low modal volumes ($V_{eff}$), and lossless coupling to the single-mode fiber with a sharp resonance peak for the resonant cavity. In this proposed structure, air-clad silica optical nanofiber is taken with a refractive index of 1.460. Two sets of mirrors are considered by maintaining a cavity length (L) as shown in fig.1 to enhance light coupling, but the scattering loss due to grating has been neglected. The proposed optimized TNFBG structure is shown in fig.1. It consists of an air clad silica ONF having diameter D = 800 nm, rectangular etching with etch width a = 119 nm, etch height b = 119 nm, etch depth c = 800 nm with grating period $\Lambda$ = 310 nm, L = 2175 nm, etched period number N = 70 and tilt angle $\theta$ = 14.5° relative to the fiber cross-section plane.

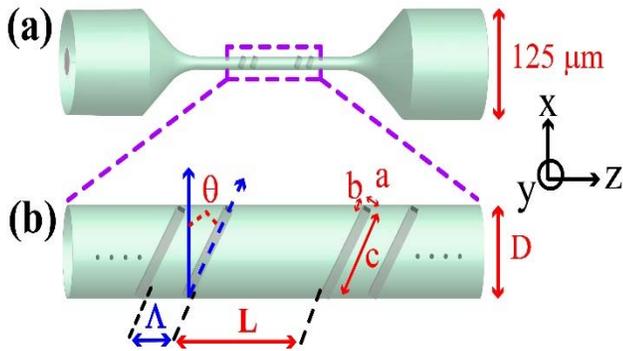

**Fig. 1.** The schematic of TNFBG. (a) Schematic diagram of an SMF of diameter 125 μm tapered to nano range with tilted grating on both sides of a cavity. (b) Enlarge View of the main structure having diameter- D, etch width- a, etch height- b, etch depth -c, cavity length- L. All gratings tilted by an angle $\theta$ having etch number N = 35 on each side of the L. Note that the number of periods in the schematic diagram is less than that in the actual simulated geometry.

To analyse the proposed system, light from a broadband source is considered to be coupled to the NFBG as shown in Fig. 2. To make the NFBG a resonant cavity, an optimized value of L = 2.175 μm is considered in the middle of two symmetric tilted gratings having a period numbers N = 35 on each side of the cavity. There is a sharp resonat peak in the middle of the stop band of the transmission spectra in both non-tiled and tilted cases. But, the resonat peak in tiled grating is more prominent and sharp as compare to non-tilted grating. Transmittance is almost double in the tilted grating than that of non-tiled grating. To have a clear modal picture over broad wavelength range, Fig. 3 shows other cavity modes that are excited on either side of the stopband. There is a stop band in the middle of the transmission spectra in a TNFBG having no cavity at the middle. We found that the depth of the wavelength band increases with the etch area (a×b) and almost 98% of the input light is reflected for N = 70. Here, the cavity wavelength can be tailored and controlled by changing the grating period $\Lambda$ and it shows a redshift as $\Lambda$ increases. The dissipative nature of resonance mode makes it different from other modes. The resonance wavelength ($\lambda_r$) of the cavity mode is 797.711 nm which corresponds to a frequency of 375.816 THz having full-width at half-maxima (FWHM) as 0.267 THz and Q factor is as high as 1408.71 due to large index modulation by tilted Bragg grating.

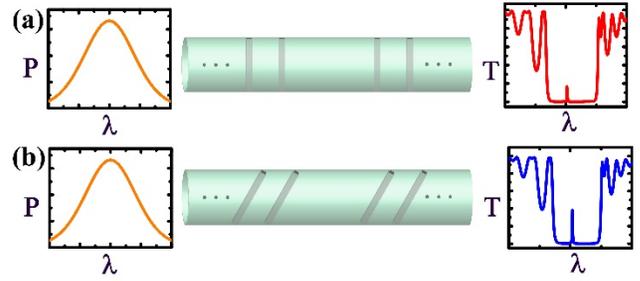

**Fig. 2.** Schematic diagram and optical response of NFBG with input Gaussian mode light (yellow curve) having power P with output transmission T spectra of (a) non-tilted grating with cavity (red curve) (b) tilted grating with a cavity (blue curve) in the centre of the tilted structure.

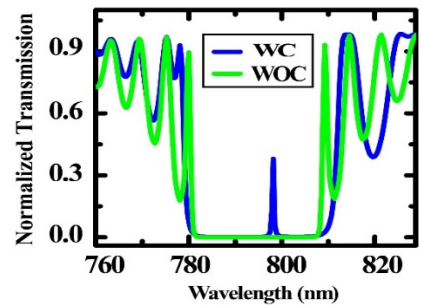

**Fig. 3.** Normalized transmission of the TNFBG without cavity (WOC)-(green curve) and with a cavity (WC) - (blue curve). The modes at higher wavelength show larger red shift in wavelength with cavity.

We found that the transmittance increases by increasing the tilt angle, and the cavity length as in fig. 4, and hence it can be tailored. Also, the propagation of the mode inside the TNFBG modulates the electric field intensity due to the presence of Bragg mirrors, a signature of Fabry-Perot (FP) like cavity modes. Figure 5(a) shows the cross-section of the normalized electric field distribution of the mode profile along the TNFBG structure at $\lambda_r$ = 797.711 nm. It can be clearly seen that the field distribution is almost symmetric on both sides of the cavity region. The zoomed-in section shown in fig. 5(b) shows higher field enhancement in the cavity region as more power couple to the structure due to the presence of the cavity. This maximum field in the cavity will enhance the spontaneous emission rate of the QEs.

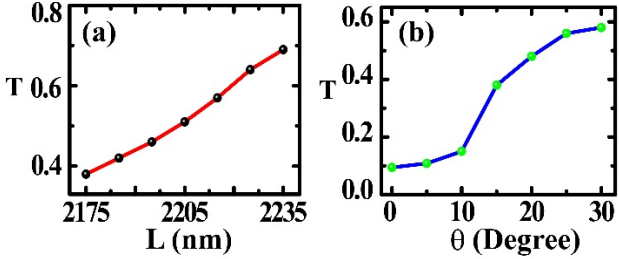

**Fig. 4.** Optical response of variation of transmittance T with (a) the cavity length L, and (b) tilt angle θ in a TNFBG

To measure the spatial confinement of the electromagnetic field in the cavity, one can calculate the effective cavity mode volume $V_{eff}$ [24],

$$V_{eff} = \frac{\int_v \varepsilon(r)|E(r)|^2 d^3r}{\max[\varepsilon(r)|E(r)|^2]} \quad (2)$$

Where E(r) is the electric field, and ε(r) is the dielectric constant of the fiber. From Eq. (2), the effective mode volume is calculated to be as small as 0.66 μm³ for TNFBG with N = 35 on each side of the cavity. However, it can be tailored with the value of N. Lower the mode volume, higher is the enhancement in the spontaneous emission decay rate.

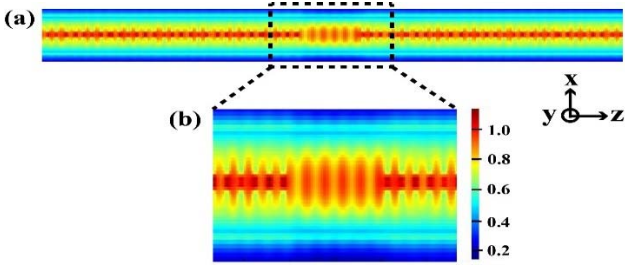

**Fig. 5.** Optical response of the nanofiber-based tilted Bragg grating with dimension (D, N, L) = (800, 35, 2175) nm with tilted angle θ = 14.5°, (a) shows the electric field distribution along the TNFBG, and (b) enlarge portion shows the higher field distribution near the cavity region.

Further, QE placed at the field antinode in an optical cavity like TNFBG will experience a medium-enhanced spontaneous emission rate relative to that in a homogenious medium based on the Purcell effect. The magnitude of enhancement of QE is given by Purcell factor F as[25],

$$F = \frac{3}{4\pi^2}\left(\frac{\lambda}{n}\right)^3 \left(\frac{Q}{V_{eff}}\right) \quad (3)$$

Where, all symbols have already been defined. This F is used as a Figure of Merit (FOM) for single photon sources. Also, Fermi's golden rule dicates that the transition rate for the atom-vacuum (or atom-cavity) system is proportional to the initial and final photon states of the system with a weak perturbation as well as the local density of state (LDOS) as,

$$\Gamma_{i \rightarrow f} = \frac{2\pi}{\hbar}|\langle f|H'|i\rangle|^2 \rho(E_f) \quad (4)$$

Where, $i$ and $f$ are the initial and final states, $H'$ is perturbed Hamiltonian and $E_f$ is final energy state. In a cavity at resonance, the density of final state is enhanced. Therefore, the spontaneous decay rate (γ) of a two level QE in the presence of the TNFBG is given by [26],

$$\gamma = \frac{\pi\omega_0}{3\hbar\varepsilon_0}|p|^2 \rho_p(r_0,\omega_0)$$

$$\rho_p(r_0,\omega_0) = \frac{6\omega_0}{\pi c^2}\left[n_p \cdot \text{Im}\{\vec{G}(r_0,r_0;\omega_0)\} \cdot n_p\right] \quad (5)$$

Where $\omega_0$ is the transition angular frequency, |p|² is the transition dipole moment, p = |p|$n_p$ is the dipole moment of an emitter, and $n_p$ is the unit vector in the direction of p. $\vec{G}(r_0,r_0;\omega_0)$ is the dydics Green's function is defined by the electric field at a point $r$ generated by a point source at point $r_0$ with dipole moment $p$ as

$$G(r,r_0) = \frac{\varepsilon_0 \varepsilon_r c^2}{p\omega_0^2} E(r) \quad (6)$$

Where, $\varepsilon_0$ and $\varepsilon_r$ are the permittivity of air and relative permittivity of medium. The enhancement factor (Γ) of QE is the power couple to the guided modes of the fiber from the emitter. The amount of coupling of lights from QE with TNFBG structure is calculated by coupling efficiency (β)- the ratio between spontaneous emission decay rate from QE with the guided mode of TNFBG to the total decay rate in all radiative direction. So,

$$\beta = \frac{\gamma_g}{\gamma_T}; \Gamma = \frac{\gamma_g}{\gamma_0}; F = \frac{\Gamma}{\beta}; F = \frac{\gamma_T}{\gamma_0} = \frac{P_T}{P_0} \quad (7)$$

Where, $\gamma_g$ is the emission rate of QE in the guided mode of TNFBG, $\gamma_T$ is the total decay rate in all radiative channel, and $\gamma_0$ is emission decay rate in free space. Also, $P_T$ and $P_0$ is the emitted power from a QE to inside the cavity and to free space respectively.

To study the coupling mechanism, fig. 5(a) shows the three different locations of placing the QE on the TNFBG. For this analysis, the x-polarized mode is considered due to its higher Q factor. Figure 5(b) shows the variation of *F* when the QE is placed straight along the nanofiber at the centre (red sphere with line) as in [27], at the surface of the nanofiber (blue sphere with line), and on the surface near the groove of the etched region (green sphere with line). As can be clearly seen, *F* has an oscillating behavior with a period of 300 nm on both sides of the zero position of the TNFBG. When the QE is placed at 150 nm away from zero on the z-axis, F increases because of increase in the electric field intensity at that point, as shown in fig. 4. It shows maximum F at the center, which is 9- fold more than at the surface and 12-fold more than near the groove's surface.

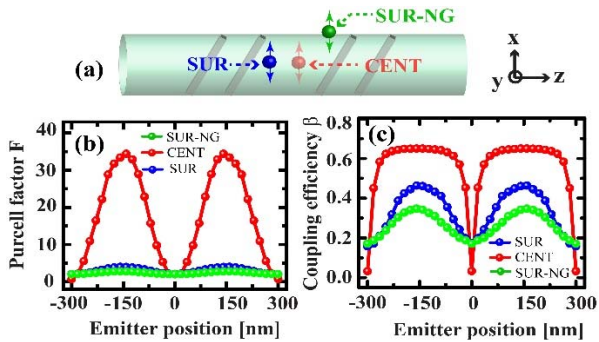

**Fig. 6.** (a) Schematic diagram of different location of QE on the TNFBG structure present at the centre (CENT)- (blurred red sphere with arrow), at the surface (SUR)- (blue sphere with arrow), and surface near groove (SUR-NG)- (green sphere with arrow). (b) Purcell factor F and (c) coupling efficiency β for different emitter locations on the extending detections of TNFBG.

Further, fig. 5(c) shows the variation of β with QE positions. It shows the maximum β at the center than at the surface. β exhibited long flat peaks spanning lengths of approximately 100 nm, followed by a sharp dip for the center position. This is because the photons emitted from the dipole selectively coupled to the fiber's fundamental mode and as a result β reaches saturation and the peak flatten. For the surface position, β shows a clear peak because the enhancement of the photon emission from the dipole is relatively small.

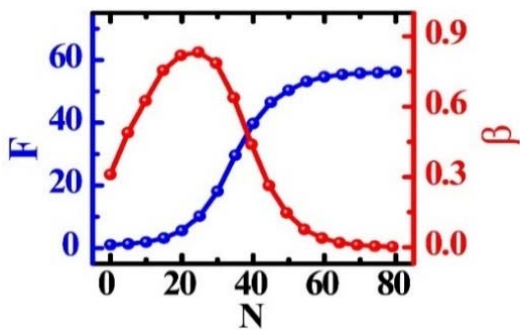

**Fig. 7.** Dependence of Purcell factor (blue) F (red) and coupling efficiency β on the period number N = *0* to 80. Note that this period number refers to one side of the TNFBG.

Further, we found that the electric field's optimal enhancement is achieved for N = 35 and beyond that electric field intensity decreases due to increase in grating loss. Further, to understand the dependence of *F* and *β* on N, fig. 6 shows that F can reach 60 for N = 80, and exhibits an increasing trend up to N = 60, and after that, it becomes flat due to the saturation of the QE's spontaneous decay rate. For N = 0, β ~ 33 % with F ~ 0.2 and for N=25, β ~ 87%. However, due to increase in period number, the grating loss increases and can be address by using lossless fiber grating.

In summary, we have reported a new avenue for quantum photonics, and cavity QED applications with F= 35 and β= 65 % for TNFBG tilt angle (θ) = 14.5º. We have shown that the electric field enhancement in the cavity region and coupling of the florescent photons from the QE at different positions in the guided mode of TNFBG. This structure would help to realize quantum information devices, such as quantum memories, fiber-integrated single-photon sources, & light-matter interaction for future quantum technologies.